\documentclass[letterpaper,notoc]{CECS}

\title{Supersymmetric extension of the nine-dimensional continuation
of the Euler density with $\mathcal{N}=2$. }
\author{Mokhtar Hassa\"{i}ne$^{1}$, Rodrigo Olea$^{2}$ and Ricardo Troncoso$^{1}$ \\
{\small {$^{1}$Centro de Estudios Cient\'{i}ficos (CECS),Casilla 1469, Valdivia, Chile.} }\\
{\small {$^{2}$Departamento de F\'{\i}sica, P. Universidad Cat\'{o}lica de
Chile,Casilla 306, Santiago 22, Chile.}}}
\preprint{{\tiny CECS-PHY-02-06} }
\abstract{
A local supersymmetric extension with $\mathcal{N}=2$ of the dimensional
continuation of the Euler-Gauss-Bonnet density from eight to nine dimensions
is constructed. The gravitational sector is invariant under local Poincar%
\'{e} translations, and the full field content is given by the vielbein, the
spin connection, a complex gravitino, and an Abelian one-form. The local
symmetry group is shown to be super Poincar\'{e} with $\mathcal{N}=2$ and a $%
U(1)$ central extension, and the full supersymmetric Lagrangian can be
written as a Chern-Simons form.}

\begin{document}

\section{Introduction}

The existence of new Lagrangians for supergravity in dimensions lower than
eleven, which cannot be obtained by dimensional reduction from the standard
Cremmer-Julia-Scherk theory \cite{C-J-S}, suggests that M-Theory may have
new ``cusps'' from which these new theories would be obtained (see e.g. \cite
{de Witt-Nicolai-Samtleben}). In this sense, exploring new dynamic and
geometric structures with local supersymmetry in dimensions $d<11$ deserves
some attention. In this spirit, we construct a local $\mathcal{N}=2$
supersymmetric extension of the Poincar\'{e} invariant gravity in nine
dimensions, possessing a different structure than the supergravity theory
obtained by dimensional reduction from standard eleven-dimensional
supergravity (see e.g. \cite{Salam-Sezgin}). The gravitational sector of
standard $\mathcal{N}=2$ supergravity in nine dimensions is described by the
Einstein-Hilbert action, and its full field content is given by $(e_{\mu
}^{a},2B_{\mu \nu },A_{\mu \nu \rho },3\phi ;2\psi _{\mu },4\chi )$ where
the scalar fields parametrize the coset $GL(2,R)/SO(2)$. In our case, the
Lagrangian for the gravitational sector is given by the dimensional
continuation of the Euler-Gauss-Bonnet density form eight to nine
dimensions, which still gives second order field equations for the metric.
The field content of this theory is the set $(e_{\mu }^{a},\omega _{\mu
}^{ab},\psi _{\mu },c_{\mu })$, where $\psi _{\mu }$ is a complex gravitino,
and $c_{\mu }$ is an Abelian one-form. Here $e_{\mu }^{a}$ is the vielbein,
and the spin connection $\omega _{\mu }^{ab}$ is now regarded as an
independent dynamical variable. The local supersymmetry algebra closes
off-shell, and is given by the super Poincar\'{e} group with a $U(1)$
central extension. The field content of the theory, then can be regarded as
the different components of a single connection for this group, and hence,
the local symmetries can be read off from a gauge transformation. This fact,
together with the existence of a fifth-rank invariant tensor for the
supergroup, allows to write the Lagrangian as a Chern density. Note that for
Chern-Simons theories the number of bosonic and fermionic degrees of freedom
do not necessarily match, since there exists an alternative to the
introduction of auxiliary fields (see e.g. \cite
{Howe-Izquierdo-Papadopoulos-Townsend}). Indeed, the matching may not occur
when the dynamical fields are assumed to belong to a connection instead of a
multiplet for the supergroup \cite{TZ-Bariloche}. For a vanishing
cosmological constant, supergravity theories sharing these last features
have been constructed in three \cite{Achúcarro-Townsend-Poincaré}, \cite
{Howe-Izquierdo-Papadopoulos-Townsend} and higher odd dimensions \cite
{BTZ,HTZ}. There exist also supergravity models with $\Lambda <0$ in three 
\cite{Achúcarro-Townsend-AdS}, five \cite{Chamseddine} and higher odd
dimensions \cite{TZ-7-11}, \cite{TZ-QMFS}.

In the next section the gravitational sector of this theory is discussed,
and its supersymmetric extension is given in Section III. Section IV is
devoted to the discussion.

\section{The gravitational sector}

As it is well-known in three dimensions, in the absence of cosmological
constant, it is possible to ensure the off-shell closure of the superalgebra
without using auxiliary fields, by demanding invariance under local
translations in tangent space. This is due to the fact that the
three-dimensional Einstein-Hilbert action is invariant under the following
transformations 
\begin{equation}
\begin{array}{lll}
\delta e^{a}=D\lambda ^{a}=d\lambda ^{a}+\omega _{\;b}^{a}\lambda ^{b} & 
;\;\; & \delta \omega ^{ab}=0\;,
\end{array}
\label{Translations}
\end{equation}
without imposing the vanishing torsion condition. In higher dimensions,
these transformations are no longer a symmetry of the Einstein-Hilbert
action. However, this symmetry is present in odd-dimensional spacetimes if
one considers an action linear in the vielbein instead of the curvature. In
particular, in nine dimensions the action reads\footnote{%
Here $e^{a}=e_{\mu }^{a}dx^{\mu }$ is the vielbein, and $R^{ab}=d\omega
^{ab}+\omega _{\;c}^{a}\omega ^{cb}$ stands for the curvature two-form.
Wedge product between forms is assumed throughout.} 
\begin{equation}
I_{G}=\int \epsilon _{a_{1}\cdots a_{9}}R^{a_{1}a_{2}}\cdots
R^{a_{7}a_{8}}e^{a_{9}}\;.  \label{Poincaregravity}
\end{equation}
Here, the Lagrangian is the dimensional continuation of the
Euler-Gauss-Bonnet density from eight to nine dimensions. This action is
singled out as the most general gravity theory, constructed out of the
vielbein and the curvature, still leading to second order field equations
for the metric, that possesses local invariance under the Poincar\'{e} group 
\cite{TZ-Grav}. Note that the transformations (\ref{Translations}) can be
read off from a gauge transformation with parameter $\lambda =\lambda
^{a}P_{a}$, assuming that the vielbein and the spin connection belong to a
connection of the Poincar\'{e} group, i.e., $A=1/2\,\omega
^{ab}J_{ab}+e^{a}P_{a}$.

Invariance under local translations (\ref{Translations}) presents an
advantage when one deals with the locally supersymmetric extension in nine
dimensions, because the closure of the superalgebra can be attained without
the need of auxiliary fields as it is discussed in the next section.

\section{Local supersymmetric extension}

The local supersymmetric extension of the gravitational action $I_{G}$ in
Eq. (\ref{Poincaregravity}) is given by the sum of three pieces

\begin{equation}
I_{9}=I_{G}+I_{\psi }+I_{c}\;,  \label{susyaction}
\end{equation}
where the fermionic term $I_{\psi }$ reads 
\[
I_{\psi }=\int \frac{i}{3}R_{abc}\bar{\psi}\Gamma ^{abc}D\psi +12\left(
R_{ab}R^{ab}R_{cd}+4(R^{3})_{cd}\right) \bar{\psi}\Gamma ^{cd}D\psi +%
\mbox{h.c.} 
\]
and the bosonic term needed to close supersymmetry $I_{c}$ is 
\begin{equation}
I_{c}=-12\int \left[
(R_{ab}R^{ab})^{2}-4R_{b}^{a}R_{c}^{b}R_{d}^{c}R_{a}^{d}\right] c\;.
\end{equation}
Here $D\psi =d\psi +\frac{1}{4}\omega ^{ab}\Gamma _{ab}\psi $ is the Lorentz
covariant derivative, and as a shorthand we have defined $%
(R^{3})_{b}^{a}=R_{c}^{a}R_{d}^{c}R_{b}^{d}$, and $R_{abc}=\epsilon
_{abc\,a_{1}\cdots a_{6}}R^{a_{1}a_{2}}R^{a_{3}a_{4}}R^{a_{5}a_{6}}$.

The action (\ref{susyaction}) is invariant under the following local
supersymmetry transformations

\begin{equation}
\begin{array}{lll}
\delta e^{a}=-i\left( \bar{\epsilon}\Gamma ^{a}\psi -\bar{\psi}\Gamma
^{a}\epsilon \right) & ; & \delta \psi =D\epsilon \\ 
\delta c=\left( \bar{\epsilon}\psi -\bar{\psi}\epsilon \right) & ; & \delta 
\bar{\psi}=D\bar{\epsilon}
\end{array}
\label{susytransformations}
\end{equation}
together with $\delta \omega ^{ab}=0$.

Note that, apart from diffeomorphisms, local Lorentz rotations, and local
Poincar\'{e} transformations (\ref{Translations}), this action possesses an
extra local $U(1)$ symmetry whose only nonvanishing transformation reads 
\begin{equation}
\delta c=du.  \label{u1symmetry}
\end{equation}

The full set of symmetries closes off-shell without requiring the
introduction of auxiliary fields. Indeed, the commutator of two
supersymmetry transformations acting on the fields is given by 
\[
\left[ \delta _{\epsilon },\delta _{\eta }\right] \left( 
\begin{array}{l}
e^{a} \\ 
\omega ^{ab} \\ 
c \\ 
\psi
\end{array}
\right) =\left( 
\begin{array}{l}
D\lambda ^{a} \\ 
0 \\ 
du \\ 
0
\end{array}
\right) \;, 
\]
where the parameter of the local translations is $\lambda ^{a}=-i(\bar{%
\epsilon}\Gamma ^{a}\eta -\bar{\eta}\Gamma ^{a}\epsilon )$ while the
parameter of the $U(1)$ symmetry is $u=\bar{\epsilon}\eta -\bar{\eta}%
\epsilon $. This means that the local symmetry group is spanned by the $%
\mathcal{N}=2$ super Poincar\'{e} algebra with a $U(1)$ central extension
whose generators are given by the set $G_{A}=\{J_{ab},P_{a},Z,Q^{\alpha },%
\bar{Q}_{\alpha }\}$. Here $J_{ab}$ and $P_{a}$ generate the Poincar\'{e}
group, $Z$ is the $U(1)\,$ central charge, and $Q^{\alpha }$, $\bar{Q}%
_{\alpha }$ are complex fermionic generators whose anticommutator reads 
\begin{equation}
\left\{ Q^{\alpha },\bar{Q}_{\beta }\right\} =-i(\Gamma ^{a})_{\beta
}^{\alpha }P_{a}+\delta _{\beta }^{\alpha }Z\;.  \label{Algebra}
\end{equation}

Let us point out that, assuming the dynamical fields to belong to a single
connection for the supergroup, i.e.,

\begin{equation}
A=\frac{1}{2}\omega ^{ab}J_{ab}+e^{a}P_{a}+cZ+\bar{\psi}Q-\bar{Q}\psi \;,
\label{A}
\end{equation}
all the local symmetries can be viewed as a gauge transformation $\delta
_{\lambda }A=d\lambda +[A,\lambda ]$, where $\lambda $ is a Lie
algebra-valued parameter. Indeed, the supersymmetry transformations (\ref
{susytransformations}) are obtained from a gauge transformation with
parameter $\lambda =\bar{\epsilon}Q-\bar{Q}\epsilon $. Analogously, the
local Lorentz transformations, local translations (\ref{Translations}), and
the local $U(1)$ symmetry (\ref{u1symmetry}) are obtained for $\lambda =%
\frac{1}{2}\lambda ^{ab}J_{ab}$, $\lambda =\lambda ^{a}P_{a}$ and $\lambda
=uZ$, respectively.

It is worth mentioning that the Lagrangian (\ref{susyaction}) can be written
as a Chern density, since it satisfies $dL_{9}=\left\langle
F^{5}\right\rangle $, where $F=dA+A^{2}$ is the curvature two-form which in
components reads

\begin{equation}
F=\frac{1}{2}R^{ab}J_{ab}+(T^{a}+i\bar{\psi}\Gamma ^{a}\psi )P_{a}+(dc-\bar{%
\psi}\psi )Z+D\bar{\psi}Q-\bar{Q}D\psi \;.  \label{Curvature}
\end{equation}
Here the bracket $\left\langle \cdot \cdot \cdot \right\rangle $ stands for
an invariant multilinear form for the super Poincar\'{e} group with $%
\mathcal{N}=2$ with a $U(1)$ central extension, whose only nonvanishing
components are given by \cite{Invariant Tensor} 
\begin{eqnarray}
\left\langle J_{ab},J_{cd},J_{ef},J_{gh},P_{i}\right\rangle  &=&\frac{16}{5}%
\epsilon _{abcdefghi}\;,  \nonumber \\
\left\langle J_{ab},J_{cd},J^{ef},J^{gh},Z\right\rangle  &=&-\frac{48}{5}%
\left[ \delta _{abcd}^{efgh}-\delta _{ab}^{ef}\delta _{cd}^{gh}\right] \;,
\label{InvariantTensor} \\
\left\langle Q^{\alpha },J^{ab},J_{cd},J_{ef},\bar{Q}_{\beta }\right\rangle 
&=&-\frac{1}{5}(\Gamma _{\;\;\;cdef}^{ab})_{\beta }^{\alpha }-\frac{3}{10}%
\delta _{cdef}^{abgh}(\Gamma _{gh})_{\beta }^{\alpha }+\frac{3}{10}\delta
_{cd}^{ab}\delta _{ef}^{gh}(\Gamma _{gh})_{\beta }^{\alpha }\;,  \nonumber
\end{eqnarray}
where (anti)symmetrization under permutation of each pair of generators is
understood when all the indices are lowered. As a consequence, the action
constructed here can be seen as a gauge theory with fiber bundle structure.
The field equations can be written in a manifestly covariant form as

\[
\left\langle F^{4}G_{A}\right\rangle =0\;, 
\]
where $G_{A}$ are the generators of the gauge group. As a direct result, the
consistency of the fermionic equations reproduces the bosonic field
equations associated to the generators appearing in the r.h.s. of the
anticommutator in Eq.(\ref{Algebra}) without imposing additional constraints 
\cite{TZ-Bariloche}. In this case, these equations correspond to the
variation with respect to the fields $e^{a}\,$and $c$.

In the Appendix B, it is shown that, if one assumes the gauge group to be
super-Poincar\'{e} admitting at most a $U(1)$ central charge, theories
featuring the properties considered here exist only in $3,5$ and $9$
dimensions.

\section{Discussion}

Here, it has been shown that the coupling of a spin $3/2$ particle to the
graviton can be achieved consitently with local superymmetry, when the
gravitational sector is described by a Lagrangian that is linear in the
vielbein rather than in the curvature, yielding second order field equations
for the metric. This corresponds to a new $\mathcal{N}=2$ local
supersymmetric extension of Poincar\'{e} invariant gravity in nine
dimensions which can be formulated as a gauge theory with a fiber bundle
structure as it occurs for Yang-Mills, and therefore it features some of the
formal advantages of the three dimensional gravity \cite
{Achúcarro-Townsend-AdS}, \cite{Witten}.

The field equations of this theory admit a class of vacuum solutions of the
form $S^{8-d}\times X_{d+1}$ where $X_{d+1}$ is a warped product of $\Bbb{R}$
with a $d$-dimensional spacetime. For this class of geometries, a nontrivial
propagator for the graviton exists only for $d=4$ and for a positive
constant cosmological \cite{HTZ}.

In nine dimensions, there exists another $\mathcal{N}=2$ theory having a
similar structure \cite{BTZ}. The corresponding gauge group is a
supersymmetric extension of Poincar\'{e} containing a fifth-rank
antisymmetric generator, so that the anticommutator of the fermionic
generators reads 
\begin{equation}
\{Q^{\alpha },\bar{Q}_{\beta }\}=-i(\Gamma ^{a})_{\beta }^{\alpha
}P_{a}-i(\Gamma ^{abcde})_{\beta }^{\alpha }Z_{abcde}\;.  \label{supertrans}
\end{equation}
Apart from the vielbein, the spin connection and a complex gravitino, a
bosonic one-form $b_{\mu }^{abcde}$ which transforms as an antisymmetric
fifth-rank tensor under local Lorentz rotations was considered \footnote{%
This kind of fields has recently become relevant in the context of dual
descriptions of linearized gravity \cite{Casini-Montemayor-Urrutia-1}, \cite
{Casini-Montemayor-Urrutia-2}, \cite{Bekaert-Boulanger}. In the last
reference a generalization of the Poincar\'{e} lemma was also provided.}. In
the model described here, the role of this field is played by the Abelian
one-form $c_{\mu }$. It would be desirable to explore whether a link between
both theories can be established.

\medskip

\textbf{Acknowledgments}

The authors are grateful to Andr\'{e}s Gomberoff and Cristi\'{a}n
Mart\'{i}nez for helpful remarks, and specially to Jorge Zanelli for
enlightening comments and for a careful reading of this manuscript. This
work is partially funded by grants 1010449, 1010450, 1020629,1040921,
3020032, 3030029 and 7010450 from FONDECYT. Institutional support to the
Centro de Estudios Cient\'{i}ficos (CECS) from Empresas CMPC is gratefully
acknowledged. CECS is a Millennium Science Institute and is funded in part
by grants from Fundaci\'{o}n Andes and the Tinker Foundation.

\section{Appendix A: Some useful formulas}

Here $\Gamma ^{a}$ stands for the Dirac matrices satisfying $\{\Gamma
^{a},\Gamma ^{b}\}=2\,\eta ^{ab}\,I$, where $a,b=1,2,\cdots ,d$, where $d$
is the spacetime dimension and $\eta ^{ab}=\mbox{diag}(-,+,\cdots ,+)$. For
the Minskowskian signature, these matrices can be chosen such that $(\Gamma
^{a})^{\dagger }=\Gamma _{0}\,\Gamma ^{a}\,\Gamma _{0}$, and the Dirac
conjugate is given by $\bar{\psi}=\psi ^{\dagger }\Gamma _{0}$. The totally
antisymmetric product of Gamma matrices is defined as 
\begin{equation}
\Gamma ^{a_{1}\cdots a_{p}}=\frac{1}{p!}\sum_{\sigma }\mbox{sgn}(\sigma
)\Gamma ^{a_{\sigma (1)}}\cdots \Gamma ^{a_{\sigma (p)}}\;.  \label{gammap}
\end{equation}
For $d=2n+1$ dimensions it is always possible to find a representation of
the Gamma matrices such that 
\[
\Gamma ^{1}\cdots \Gamma ^{d}=(-i)^{n+1}I\;, 
\]
and hence, one obtains the following relation

\begin{equation}
\Gamma ^{a_{1}\cdots a_{d-p}}=(-1)^{\frac{p(p-1)}{2}}\frac{(-i)^{n+1}}{p!}%
\,\epsilon ^{a_{1}\cdots a_{d-p}\;b_{1}\cdots b_{p}}\Gamma _{b_{1}\cdots
b_{p}}\;.  \label{gammaduality}
\end{equation}

All fermionic terms in the actions considered here take the form 
\[
I_{\psi }^{p}=\int X_{a_{1}\cdots a_{p}}\left[ \bar{\psi}\Gamma
^{a_{1}\cdots a_{p}}D\psi +\mbox{h.c.}\right] \;, 
\]
where $X_{a_{1}\cdots a_{p}}$ is a covariantly constant $(d-3)$-form
satisfying $DX_{a_{1}\cdots a_{p}}=0$, which does not transform under
supersymmetry, i.e., $\delta X_{a_{1}\cdots a_{p}}=0$. The variation of $%
I_{\psi }^{p}$ under the supersymmetry transformations (\ref
{susytransformations}) is given by 
\begin{eqnarray}
\delta I_{\psi }^{p} &=&-\frac{1}{4}\int X_{a_{1}\cdots a_{p}}R_{ab}\left[ 
\bar{\epsilon}\left\{ \Gamma ^{a_{1}\cdots a_{p}},\Gamma ^{ab}\right\} \psi -%
\mbox{h.c.}\right]  \nonumber \\
&=&-\frac{1}{2}\int X_{a_{1}\cdots a_{p}}R_{ab}\left[ \bar{\epsilon}\Gamma
^{a_{1}\cdots a_{p}ab}\psi -\mbox{h.c.}\right]  \label{Golden} \\
&&+\frac{p(p-1)}{2}\int X_{a_{1}\cdots a_{p-2}ab}R^{ab}\left[ \bar{\epsilon}%
\Gamma ^{a_{1}\cdots a_{p-2}}\psi -\mbox{h.c.}\right] \;,  \nonumber
\end{eqnarray}
up to a boundary term. In our case, $X_{a_{1}\cdots a_{p}}$ always involves
the contraction of $(n-1)$ curvatures leaving $p$ free indices. \newline

\section {Appendix B: Three, five and nine dimensions}

The only action for gravity, constructed out of the vielbein and the
curvature, leading to second order field equations for the metric, and
possessing local invariance under the Poincar\'{e} group exists only for $%
d=2n+1$ dimensions, and is given by\footnote{%
Lorentz-Chern-Simons forms, which are trivially invariant under
supersymmetry, can also be considered. However, as it occurs in three
dimensions \cite{Deser-Jackiw-Templeton}, this would yield third order field
equations for the metric in the vanishing torsion sector.}

\begin{equation}
I_{G}=\int \epsilon _{a_{1}\cdots a_{2n+1}}R^{a_{1}a_{2}}\cdots
R^{a_{2n-1}a_{2n}}e^{a_{2n+1}}=\int R_{abc}R^{ab}e^{c}\;,  \label{I2n+1}
\end{equation}
where $R_{abc}$ is a shorthand for $R_{abc}=\epsilon _{abca_{1}\cdots
a_{2n-2}}R^{a_{1}a_{2}}\cdots R^{a_{2n-3}a_{2n-2}}$.

We explore now whether this theory accepts a supersymmetric extension, for
which the dynamical fields belong to a connection of the standard
super-Poincar\'{e} group admitting at most a $U(1)$ central charge. This
means that the field content, apart from the vielbein, the spin connection
and a complex gravitino, should eventually be supplemented by a one-form $%
c_{\mu }$. In addition, the supersymmetry transformations remain the same as
in Eq.(\ref{susytransformations}), regardless the space-time dimension.

The variation of (\ref{I2n+1}) under supersymmetry is 
\begin{equation}
\delta I_{G}=-i\int R_{abc}R^{ab}\left( \bar{\epsilon}\Gamma ^{c}\psi -%
\mbox{h.c.}\right) \;,  \label{varI2n+1}
\end{equation}
which must be cancelled by the variation of a fermionic term, that is
quadratic in the gravitini without involving the vielbein or the $c$-field.
Lorentz covariance singles out this fermionic term to be 
\begin{equation}
\frac{i}{3}\int R_{abc}\left( \bar{\psi}\Gamma ^{abc}D\psi +\mbox{h.c.}%
\right) \;.  \label{fermionic3}
\end{equation}
Its variation, in turn, produces an additional contribution of the form 
\begin{equation}
-\frac{i}{6}\int R_{abc}R_{de}\left( \bar{\epsilon}\Gamma ^{abcde}\psi -%
\mbox{h.c.}\right) \;.  \label{bosonic}
\end{equation}
In three dimensions, this term identically vanishes, and hence, the
supergravity action does not require a central extension\footnote{%
Had we dealt with Majorana spinors, the same argument would have held, and
the theory for $\mathcal{N}=1$, discussed in Ref. \cite{Marcus-Schwarz} is
recovered.}. In five dimensions, since $\Gamma ^{abcde}$ is proportional to
the Levi-Civita tensor, the extra bosonic term $2\int R_{ab}R^{ab}c$, must
be necessarily added to cancel (\ref{bosonic}). In this form, the minimal
five dimensional supergravity is obtained for the $U(1)$ centrally extended
super-Poincar\'{e} group \cite{BTZ}.

In order to see whether the term (\ref{bosonic}) can be canceled in higher
odd dimensions, it is useful to express it as a linear combination of 
\begin{equation}
\int (R^{3})_{a_{1}a_{2}}R_{a_{3}a_{4}}\cdots R_{a_{d-6}a_{d-5}}\left( \bar{%
\epsilon}\Gamma ^{a_{1}\cdots a_{d-5}}\psi -\mbox{h.c.}\right) \;,
\label{bosonic1}
\end{equation}
and 
\begin{equation}
\int R^{2}R_{a_{1}a_{2}}R_{a_{3}a_{4}}\cdots R_{a_{d-6}a_{d-5}}\left( \bar{%
\epsilon}\Gamma ^{a_{1}\cdots a_{d-5}}\psi -\mbox{h.c.}\right) \;.
\label{bosonic2}
\end{equation}
In seven dimensions, even though the term (\ref{bosonic2}) can be
compensated by 
\begin{equation}
\int R^{2}\left( \bar{\psi}D\psi +\mbox{h.c.}\right) \;,  \label{Seven1}
\end{equation}
the remaining one (\ref{bosonic1}) can never be canceled. Indeed, this term
reduces to 
\begin{equation}
\int (R^{3})_{ab}\left( \bar{\epsilon}\Gamma ^{ab}\psi -\mbox{h.c.}\right)
\;,  \label{d71}
\end{equation}
and can not be eliminated by the variation of a bosonic piece. On the other
hand, by virtue of formula (\ref{Golden}), the only fermionic term,
different from (\ref{fermionic3}), whose variation is cubic in the curvature
times the combination $\bar{\epsilon}\Gamma ^{ab}\psi $ is given by (\ref
{Seven1}). However, its variation can never cancel the term in (\ref{d71}),
and hence, under our assumptions, the seven dimensional case is ruled out.

Following the same procedure in higher dimensions, invariance under local
supersymmetry would demand the introduction of a growing series of fermionic
terms in the Lagrangian of the form 
\begin{equation}
I_{\psi }^{p}=\int X_{[p]}\left[ \bar{\psi}\Gamma ^{[p]}D\psi +\mbox{h.c.}%
\right] \;,  \label{generalfermionic}
\end{equation}
where $\Gamma ^{[p]}=\Gamma ^{a_{1}\cdots a_{p}}$, and $X_{[p]}$ is a $(d-3)$%
-form constructed exclusively from curvatures $R_{ab}$ \cite{Fermionic}.
Exhausting all the relevant combinations of the fermionic terms in (\ref
{generalfermionic}) and also the bosonic ones containing the Abelian field $%
c $, it is shown that supergravity theories with local invariance under
super-Poincar\'{e} with a $U(1)$ central charge do not exist for higher odd
dimensions different from nine.

For dimensions $d=4k+1>5$, the variation (\ref{bosonic1}) can only be
canceled by 
\begin{equation}
\int (R^{3})_{a_{1}a_{2}}R_{a_{3}a_{4}}\cdots R_{a_{k-1}a_{k}}\left( \bar{%
\psi}\Gamma ^{a_{1}\cdots a_{k}}D\psi +\mbox{h.c.}\right) \;,  \label{4k+1}
\end{equation}
and following the Noether procedure, after a number of steps, a term of the
form 
\begin{equation}
\int (R^{3})_{ab}\left( R^{2k-3}\right) _{cd}\left( \bar{\epsilon}\,\Gamma
^{abcd}\psi -\mbox{h.c.}\right) \;,  \label{problematic4k+1}
\end{equation}
necessarily appears in the variation\footnote{%
Here $(R^{m})_{b}^{a}=R_{c_{1}}^{a}R_{c_{2}}^{c_{1}}\cdots R_{b}^{c_{m-1}}$.}%
. For $k>2$ this last expression can not be canceled by a bosonic term,
while using formula (\ref{Golden}), it might be compensated resorting to
terms proportional to $I_{\psi }^{6}$ and $I_{\psi }^{2}$. However, none of
the possibilities are satisfactory. In fact, in the first case, the explicit
construction of the terms along $\Gamma ^{[6]}$ leads us back exactly to the
expression that generates the term (\ref{problematic4k+1}). For the second
case, formula (\ref{Golden}) implies that the variation of any of the
possible terms contains at least one curvature that is not contracted with
another curvature, and therefore, the term (\ref{problematic4k+1}) can never
be canceled. Consequently, for $k>2$, local supersymmetry is never attained.

The nine dimensional case $(k=2)$ is exceptional because the leftover term (%
\ref{problematic4k+1}) involves a curvature that is not contracted with
another one. Unlike the previous cases, this fact allows to cancel (\ref
{problematic4k+1}) by means of the additional term 
\[
\int (R^{3})_{ab}\left( \bar{\psi}\Gamma ^{ab}D\psi +\mbox{h.c.}\right) \;, 
\]
that produces also a contribution which is compensated by a bosonic term
containing the $c$-field given by 
\[
\int (R^{3})_{ab}R^{ab}c\;. 
\]
Analogously, the remaining term in Eq.(\ref{bosonic2}) is canceled by the
variation of a fermionic term and a bosonic one depending on the Abelian
field. In this case, the full supergravity action in Eq.(\ref{susyaction})
is recovered.

For the remaining dimensions, $d=4k-1>7$, in order to cancel the variation
in Eq.(\ref{bosonic1}), after a number of steps, it is inevitable to add the
following term to the action 
\begin{equation}
\int (R^{3})_{ab}(R^{2k-5})_{cd}\left( \bar{\psi}\Gamma ^{abcd}D\psi +%
\mbox{h.c.}\right) \;,  \label{4k-1}
\end{equation}
whose variation contains the term 
\begin{equation}
\int \left( R^{2k-1}\right) _{ab}\left( \bar{\epsilon}\,\Gamma ^{ab}\psi -%
\mbox{h.c.}\right) \;.  \label{problematic4k-1}
\end{equation}
Here the argument to rule out these dimensions is similar to the one for $%
d=7 $. Indeed, this last term can not be canceled by a bosonic one, and by
virtue of formula (\ref{Golden}), the only fermionic term different from (%
\ref{4k-1}), whose variation leads to a $(2k-1)$-th power of the curvature
times the combination $\bar{\epsilon}\Gamma ^{ab}\psi $ must be of the form
given by $I_{\psi }^{0}$ in Eq.(\ref{generalfermionic}). However, as it can
be seen from formula (\ref{Golden}), their variation always involves the
combination $R_{ab}\Gamma ^{ab}$, and thus the term (\ref{problematic4k-1})
can never be canceled.

In summary, in this Appendix we have shown that \textit{the supersymmetric
extension of the action (\ref{I2n+1}), that possesses local invariance under
the super-Poincar\'{e} group admitting at most a }$U(1)$\textit{\ central
charge, exists only in }$3$\textit{, }$5$\textit{, and }$9$\textit{\
dimensions.}


\begin{thebibliography}{99}
\bibitem{Invariant Tensor}  A fifth-rank invariant tensor $\Delta _{ABCDE}$
which is (anti)symmetric under permutation of any pair of indices, depending
on the Grassmann parity of the generators, satisfies the condition 
\[
C_{BA_{1}}^{F}\Delta _{FA_{2}\cdot \cdot \cdot A_{5}}+\cdot \cdot \cdot
+C_{BA_{5}}^{F}\Delta _{A_{1}\cdot \cdot \cdot A_{4}F}=0\;,
\]
where $C_{BC}^{A}$ are the structure constants. In nine dimensions, the
existence of this invariant tensor is guaranteed as long as the central
charge $Z$ is present, and it can be written in terms of a multilinear form
of the generators as $\Delta _{ABCDE}=\left\langle
G_{A},G_{B},G_{C},G_{D},G_{E}\right\rangle $.
\bibitem{Fermionic}  Indeed, $X_{a_{1}\cdots a_{p}}$ never contains
vielbeins and it is also independent of the $c$-field. Thus, the expression (%
\ref{generalfermionic}) is the most general one which is quadratic in the
fermions and compatible with Lorentz covariance. Note that quartic terms in
the fermions are never needed.
\end{thebibliography}
\end{document}